\documentclass[twocolumn,showpacs,preprintnumbers,amsmath,amssymb]{revtex4}

\usepackage{graphicx}
\usepackage{dcolumn}
\usepackage{bm}

\begin{document}


\title{Long-range dispersion and spatial diffusion of fault waves in the Burridge-Knopoff earthquake model.} 
\author{Alain~M.~Dikand\'e}
\affiliation{D\'epartement de Physique, Facult\'e des Sciences, Universit\'e de Sherbrooke J1K2R1 Sherbrooke Qu\'ebec, Canada.}  
\email{amdikand@physique.usherb.ca}
\date{\today}

\begin{abstract}
The Burridge-Knopoff model of earthquakes has recently gained increased interest for the consistency of 
the predicted energy released by sismic faults, with the Gutenberg-Richter scaling law. The 
present work suggests an improvement of this model to account for long-range dispersions and large spatial diffusion of sismic faults. 
An enhancement of the threshold speed of shock waves driven by translated fault fronts is pointed out and shown to result from the 
interactions between components of the system situated far aways them and others. Due to the enhanced threshold speed, 
size of the sismic fault gets increased but a control effect can still be gained from tunable dispersion extent irrespective 
of the total length of the system. To the viewpoint of the Burridge-Knopoff block-lattice model, this last consideration 
introduces the possibility of sizable but finite interactions among infinitely aligned massive blocks. Implications on the 
fault wave propagation are examined by numerical simulations of the improved nonlinear partial differential equation. 
\end{abstract}

\maketitle
Understanding natural catastrophes is one of the greatest challenges faced by scientists from various fields. Recently, 
earthquake phenomena attracted a lot of attention to both viewpoints of the statistics of earthquake 
events~\cite{hallgass,sornette} and of the dynamics of sismic faults~\cite{carlson1}. The first~\cite{bak1,bak2}, based on 
cellular automata models and involving fractal faults, assumes a system which is in perpetual crisis~\cite{carlson1,bak1}. 
The main feature of this approach is the prediction of large fluctuations in the avalanche sizes~\cite{bak1}. The second 
approach follows a deterministic description assuming sismic events as time evolving processes. More explicitely, this last 
approach is based on spatio-temporal 
evolutions of the sismic fault considered within equations which in general are of the Klein-Gordon type. The best known 
of these equations is provided by the Burridge-Knopoff(BK) model~\cite{burridge} intended for two parallel tectonic 
plates subjected to stick-slip frictions, which can slide one relative to the other under a velocity-weakening driving force. 
In theory, one considers the motion of massive underground blocks attached to a static surface 
and interacting them and others via harmonic springs of constant strengths. The dynamic friction force introduces 
nonlinearity in the fault dynamics so that the corresponding discrete nonlinear equation writes~\cite{carlson1,langer}: 
\begin{equation}
M \ddot{U}_n= K(U_{n+1} - 2U_n + U_{n-1}) - k_o U_n - \Phi(\dot{U}_n/V_o),     \label{eqat1}
\end{equation}
where $U_n$ is the relative sliding amplitude of the $n^{th}$ block of mass $M$, $K$ is the effective coupling strength 
felt by the block in the presence of two nearest-neighbour blocks, $k_o$ is the uniform strength of block stick to the surface 
and dots refer to time derivatives. We assume two tectonic plates at equilibrium, in other words we neglect residual uniform sliding 
motions of blocks as they are pulled individually forward(or backward) relative to the upper plate~\cite{carlson1}. 
Concentrating ourselves on the sismicity of the model triggered  by the dynamic friction force $\Phi(\dot{U}_n/V_o)$, this 
force is usually written:
\begin{equation}
\Phi(\dot{U}_n/V_o)= F_o/(1 + \dot{U}_n/V_o),       \label{eqat2}
\end{equation}
in which $F_o$ and $V_o$ are two constant parameters. Equation~(\ref{eqat1}) exhibits rich physical properties one most 
relevant being the Gutenberg-Richter~\cite{richter} scaling law associate to the distribution of the energy released by 
earthquakes. Otherwise, previous numerical simulations~\cite{carlson3} established an agreement of the resulting 
velocity profiles and wave amplitudes with real earthquake processes and routes to chaos have very recently been investigated~\cite{souza}. \\ 

\begin{figure}
\includegraphics[height=6cm]{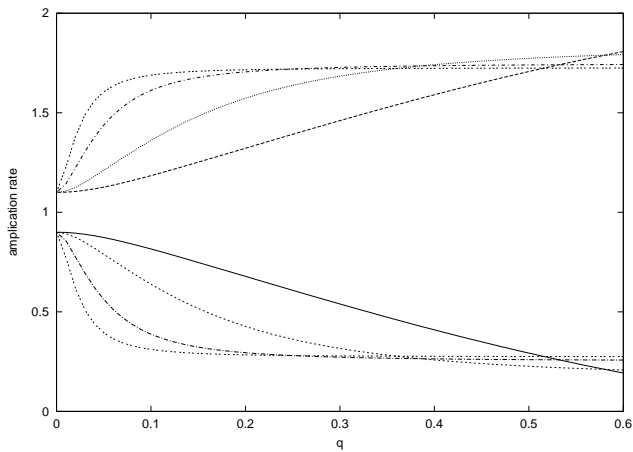}
\caption{\label{fig:spect} Dispersion laws versus reduced wavector $Q/\pi$. From the upper curve in the "high-frequency" regime(upper sub-spectrum): 
$r=$ 0.9, 0.8, 0.5, 0.}
\end{figure}

\begin{figure}
\includegraphics[height=6cm]{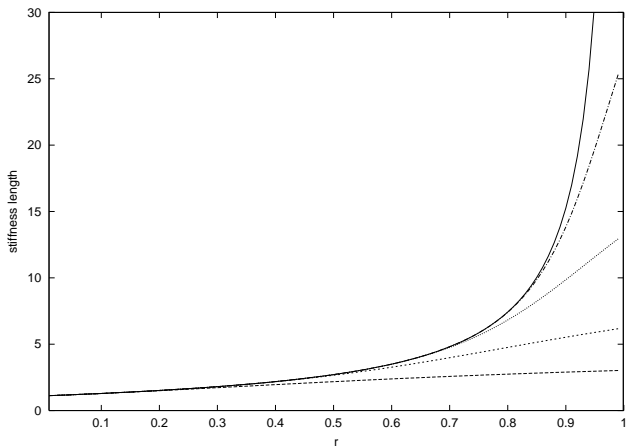}
\caption{\label{fig:vitesse} threshold speeds as a function of $r$. From the lowest curve: $L=$  4, 9, 20 41, 500.}
\end{figure}

\begin{figure*}\centering
\begin{minipage}[t]{0.5\textwidth}
\centering
\includegraphics[width=3.6in]{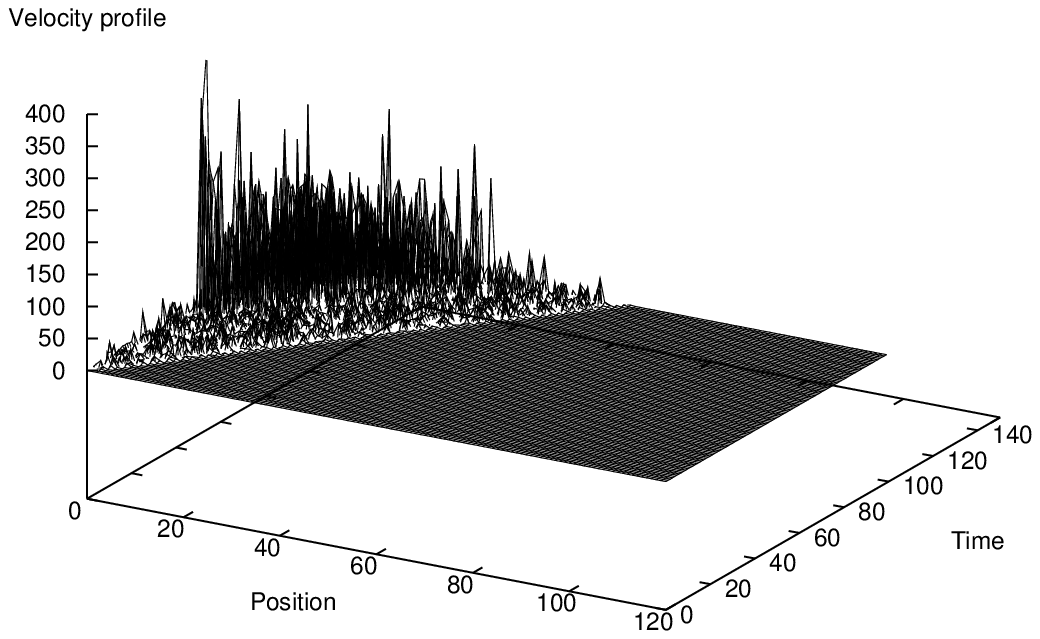}
\caption{\label{fig:sol1} Velocity profile for $r=0$. }
\end{minipage}%
\begin{minipage}[t]{0.5\textwidth}
\centering
\includegraphics[width=3.6in]{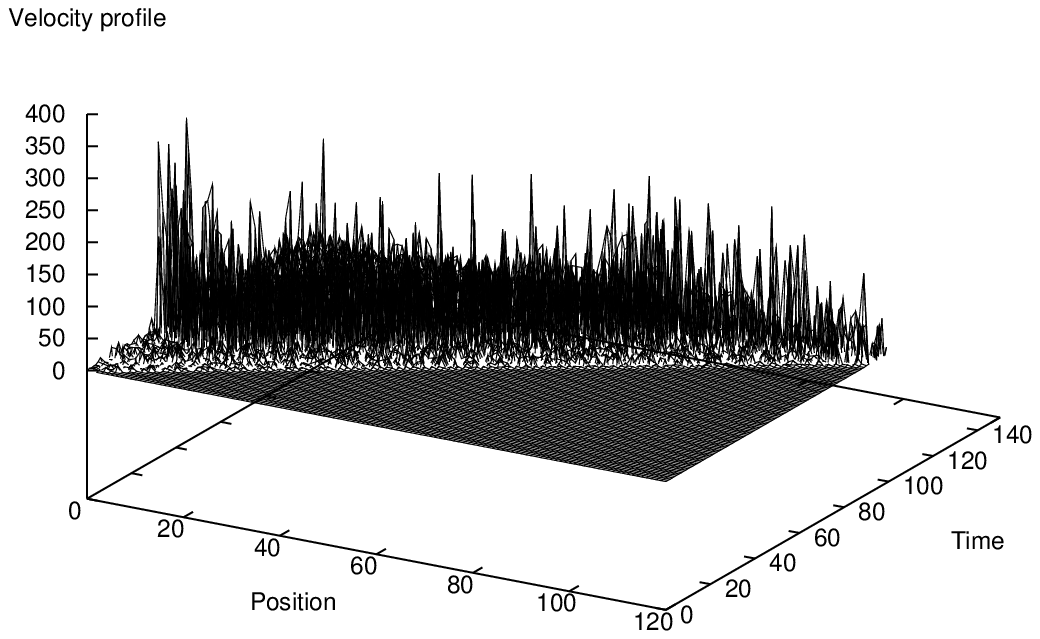}
\caption{\label{fig:sol2} Velocity profile for $r=0.2$.}
\end{minipage}
\end{figure*}

\begin{figure*}\centering
\begin{minipage}[t]{0.5\textwidth}
\centering
\includegraphics[width=3.6in]{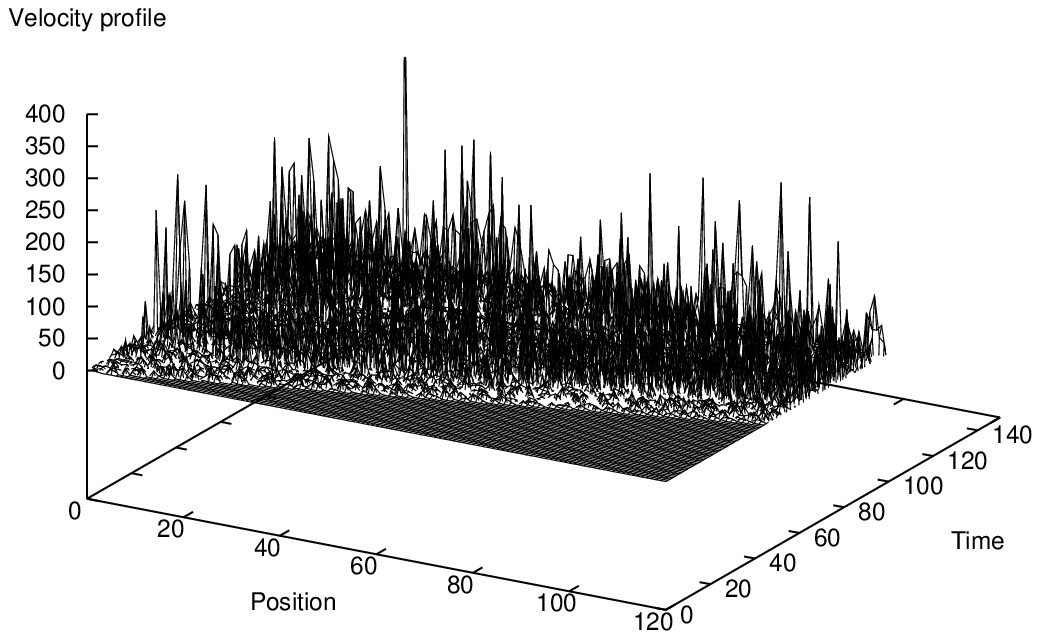}
\caption{\label{fig:sol3} Velocity profile for $r=0.4$. }
\end{minipage}%
\begin{minipage}[t]{0.5\textwidth}
\centering
\includegraphics[width=3.6in]{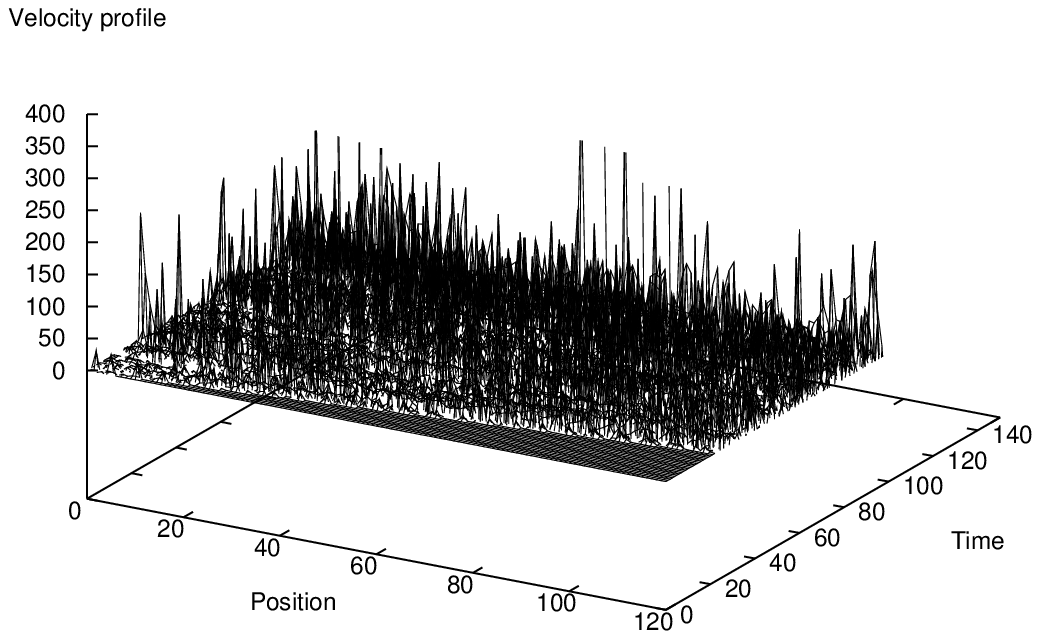}
\caption{\label{fig:sol4} Velocity profile for $r=0.6$.}
\end{minipage}
\end{figure*}

Discreteness is however another relevant aspect of this equation and constitutes the main subject of the present work. 
Discreteness connects to dispersion which is an intrinsic property of 
the system having direct incidence on its stability. Dispersion in the present system is as 
foundamental as it characterizes the ability of the interacting block lattice to respond to both small and large 
amplitude excitations propagating along the lattice, namely its fixes the threshold frequency and speed for stable 
excitations. This problem has been discussed in previous studies~\cite{carlson3,meyers} in terms of an amplification 
rate from which the group velocity was derived by linearizing this physical parameter about a characteristic wavector.
As it stands, the BK model supposes each block can only see the two nearest-neighbour blocks which means short-range 
dispersion. Accordingly, the corresponding threshold speed will be appropriate only for short-range(SR) dispersive excitations. 
However, spring models are flexible in essence and SR descriptions often consist of approximations of the actual physics behind 
these models. For instance, flexibility of the lattice of massive blocks coupled by springs 
can allow a given $n^{th}$ block to respond almost instantaneously to a perturbation taking born on or crossing an $m^{th}$ 
block relatively far away(i.e. beyond nearest neighbours) from it. Numerous sismic events show clear evidences of an 
implication of such features and it is of crucial importance to take them into account. Moreover, recent advances on 
spring models make possible their theoretical descriptions in terms of long-range(LR) dispersions~\cite{diamant} and the 
associate Klein-Gordon equation is qualitatively similar to the SR one. Nevertheless, the SR coupling strength $K$ is substituted 
for a LR dispersion potential such that the system dynamics now obeys the following equation:
\begin{equation}
M \ddot{U}_n= \sum_{m\neq n}{K(m-n)(U_m - U_n) - k_o U_n - \Phi(\dot{U}_n/V_o)},     \label{eqat3}
\end{equation}
The LR dispersion force $K(m-n)$ is thus an adjustable parameter and the equation~(\ref{eqat3}) turns to a deformable BK 
model. One most interesting features of this deformability is the possibility to adjust the model to various 
situations according to the strength and nature of the interactions among components of the system. For instance, if we 
take account of the block stick to the upper plate and assume blocks to be relatively strongly attached to the surface, 
the LR dispersion function should quickly decrease at short distances and an exponentially decreasing function with increasing 
distance $m-n$ can be an acceptable approximation. This last consideration is formulated assuming an expression 
of the form:
\begin{equation}
K(\ell)= K \frac{1-r}{r(1-r^L)}r^{\vert \ell \vert}, \hspace{.1in} \ell = m-n. \label{eqat4}
\end{equation} 
where $L$ is the spatial extent of the interaction between blocks. By 
spatial extent we understand the number of neighbour blocks from a given block which can be different from the total number of 
blocks forming the whole lattice. This precision is of particular importance here since we want a model that allows a control of the spatial extent 
of the dispersion. In addtion to $L$, the quantity $r$ is another control parameter but relates instead to the strengths of successive couplings. For an 
exponential fall-off we need $r \sim e^{-\lambda}$ where $\lambda$ is a positive constant. Or also, we can just confine 
$r$ in the interval $0 \leq r < 1$. In this way, the LR dispersion function~(\ref{eqat4}) looks quite like the so-called 
Kac-Baker potential~\cite{kac,lieb,sarker,dik}. Nevertheless, unlike the usual Kac-Baker potential the new version~(\ref{eqat4}) has finite and constant 
magnitude for finite values of $L$. Analytically, this is traduced by the constraint:           
\begin{equation}
\sum_{\ell=1}^L K(\ell)= K \label{sum}
\end{equation} 
This constraint acquires particular interest in the present context since by fixing the total interaction between blocks 
to a finite magnitude, we avoid uncontrollable dispersions of the energy carried by the LR excitations which, otherwise, 
is nothing else but the total energy carried by the fault. \\
In what follows we examine dispersion properties of the improved model~(\ref{eqat3}) and point out some 
consequences of the account of the LR dispersion on the fault dynamics. 
Following usual considerations~\cite{carlson3,meyers}, we linearize this equation assuming small-amplitude motions of 
blocks about their equilibra. These equilibra correspond to the most stable positions in the lattice after uniformly 
pulling the blocks, resulting in a uniform equilibrium position $U_o= F_o/k_o$. Setting 
$U_n(t)= U_o + u(q)\, \exp\left(n \, a \, Q i + \Omega t\right)$ and keeping only linear terms in $u$ and $\dot{u}$, 
equation~(\ref{eqat3}) leads to the following dispersion law:
\begin{eqnarray}
\Omega_{\pm}(\tilde{Q})&=& \Omega_o \left[ 1 \pm \sqrt{1 - \frac{M}{k_o} \left(1 - \frac{K_L(\tilde{Q})}{k_o} \right) \left(\frac{2V_o}{U_o}\right)^2 } \right], \nonumber \\
\Omega_o&=& \frac{F_o}{M V_o}, \hspace{.2in} \tilde{Q}= Q a. \label{dispers}
\end{eqnarray}
$\Omega$ in this last relation is the amplification rate for small-amplitude excitations and $Q$ harmonics associate to their spatial dispersions. 
The function $K_L(\tilde{Q})$ appearing in~(\ref{dispers}) is defined as:
\begin{equation}
K_L(\tilde{Q})= 2 \sum_{\ell=1}^L{K(\ell)\left[\cos(\tilde{Q} \ell ) -  1\right]}, \label{disp}
\end{equation}
where $a$ is the equilibrium separation between two neighbouring blocks. Thus, $a$ appears as a characteristic length scale of the 
model. As we will see, the function $K_L(\tilde{Q})$ governing spatial dispersion provides another 
relevant length scale. Instructively, this second length scale was previously considered and called stiffness length(e.g. denoted $\xi$ in~\cite{langer}). 
Below we keep the same viewpoint but introduce more suggestive interpretation in terms of the size of fault wave as it moves in a Galilean frame 
i.e. with a translated wavefront. To start let us examine the dispersion law obtained in~(\ref{dispers}). The 
discrete sum in $K(\tilde{Q})$ can be calculated analytically using the identity:
\begin{eqnarray}
 V_L(\tilde{Q})&=& \sum_{\ell=1}^L r^{\ell} \cos(\tilde{Q} \ell)\nonumber \\
      &=& \frac{r \cos(\tilde{Q}) - r^2 - r^{(L+1)} [\cos((L+1)\tilde{Q}) - r \cos(L\tilde{Q}) ]}{1 - 2r \cos(\tilde{Q}) + r^2}. \nonumber \\
                                                     \label{disper1}
\end{eqnarray}
With help of the analytical expression of $\Omega(\tilde{Q})$ derived from this identity an explicit formulation of the dispersion 
property of the LR system becomes trivial. In this goal, note the presence of two signs $\pm$ in~(\ref{dispers}) which relate to shock fronts 
moving(translating) respectively to the left(backward) and to the right(forward) in the block lattice. To clearly see these two distinct polarities in the dispersion 
properties of the fault wave, on figure~\ref{fig:spect} we plot the two amplification rates $\Omega_{\pm}$  as function of the wavector $Q/\pi$ for 
the infinite-extent dispersion and for some arbitrary values of the $LR$ parameter $r$. The zero dispersion mode $Q=0$ appears as lying inside a gap 
of finite width $\Delta \Omega= \vert\Omega_+ - \Omega_- \vert_{Q\rightarrow 0}$. The effect of this gap on the overall lattice dispersion 
is to lift the degeneracy of the dispersion spectrum giving rise to two separate sub-spectra associate to backward 
and forward dispersion modes. However, both sub-spectra possess common characteristic(sound) speed which is approximately the 
slope of the linear part of their dispersion curves and consequently can be estimated from:
\begin{equation}
C_L= (1/M) \partial_Q \Omega_{\pm}(Q) \vert_{Q \rightarrow 0}, \label{vite}
\end{equation}
$C_L$ is plotted on fig.~\ref{fig:vitesse} as a function of $r$ assuming different values $L$. A stricking feature in this last 
figure is an enhancement of the threshold speed as the number of interacting neighbours increases. 
The infinite increase of the characteristic speed for infinite interaction extent indicates a system becoming strongly 
unstable as all components forming this system are almost sensitives to their mutual presence at relatively long 
distances them and others. \\
Implications of this infinite increase of the characteristic speed become quite clear if we consider excitations with 
long wavelengths compared to the characteristic length scale $a$, then readily assumed as an ultra-violet cut-off. These are 
continuum-limit processes of the equation~(\ref{eqat3}) and hence can be described by the continuum nonlinear Klein-Gordon 
equation: 
\begin{equation}
\ddot{U}_n = C_L^2 U_{xx} - \omega_o^2 U - \frac{1}{M} \Phi(\dot{U}_n/V_o),     \hspace{.1in} \omega_o^2= \frac{k_o}{M}.   \label{eqat5}           
\end{equation}
where $n\, a \rightarrow x$. To see the meaning of $C_L$ to the system dynamics, we perform Galilean transformation 
$U(s, t) \rightarrow U(s \pm vt)$ setting: $z= \frac{x \pm vt}{\gamma \xi_L}$, $\gamma^2= 1 - \upsilon^2/C_L^2$ and find:
\begin{eqnarray}
Y_{zz} + Y - 1/\left(1 \pm \alpha Y_z\right) =0,  \hspace{.1in} \xi_L^2 = \frac{C_L^2}{\omega_o^2}, \hspace{.1in} \alpha= \frac{\upsilon U_o}{\xi_L V_o}, \label{eqat6}           
\end{eqnarray}
with $Y = U/U_o$. Eq.~(\ref{eqat6}) reveals that $\xi_L$ is precisely the characteristic topological 
size of the sismic fault, hence a second characteristic length scale of the model. The linear dependence of this quantity on 
$C_L$ is suggestive enough as for the effect of the LR dispersion on this second length scale.  \\
  To end, let us look at the consequence of an account of the LR dispersion on the spatial diffusion of large-amplitude(soliton-like) 
fault waves. For this purpose, the continuum nonlinear equation~(\ref{eqat6}) was solved(in $U$) approximating the diffusion 
operator(second-order spatial derivatives) within a central-difference scheme and applying combined fourth-order Runge-Kutta and 
fifth-order Runge-Kutta rules(so-called Runge-Kutta-Fehlberg algorithm) with error controls on time-dependent variables. 
Arbitrary values of the model parameters were chosen except $\xi_L$ whose dependence on the two control parameters $r$ 
and $L$ were carefully treated. To this last point, according to fig.~\ref{fig:vitesse} the effect of increasing the dispersion 
extent $L$ keeping $r$ fixed is qualitatively similar to the effect of an increase of $r$ for a 
fixed value of $L$. Therefore results displayed below assume two different but equivalent qualitative interpretations. On 
figs.~\ref{fig:sol1},~\ref{fig:sol2},~\ref{fig:sol3} and~\ref{fig:sol4}, we present some velocity profiles as the topological size $\xi_L$ is 
increased by tuning either $r$ fixing $L$, or $L$ fixing $r$. Initial conditions are $(V, \dot{V})= (-1.0, 0)$. 
The velocity profile in fig.~\ref{fig:sol1} can serve as a reference since it corresponds to the SR 
solution($r=0$, $L \rightarrow \infty$). The three next figures 
are spatio-temporal shapes of the velocity profile for $r=0.2$ 0.4 and 0.6 respectively(with $L=40$ nearest 
neighbours). As one notes, the wave carrying the sismic fault gradually spreads all over the space as $r$ increases. 
This suggests an increasingly large diffusion of the fault due to a growth of the topological size of this wave as $r$ increases. 
Therefore, numerical simulations are in excellent agreement with our analytical predictions.    
\begin{acknowledgments}
Enriching discussions with T.~C.~Kofan\'e are greatly appreciated.
\end{acknowledgments}
 

\begin{thebibliography}{}\label{sec:references}
\bibitem{hallgass} R. Hallgass, V. Loreto, O. Mazzella, G. Paladin and L. Pietronero, Phys. Rev. E56, 1346(1997).
\bibitem{sornette} M. W. Lee, D. Sornette and L. Knopoff, Phys. Rev. Lett. 83, 4219(1999).
\bibitem{carlson1} J. M. Carlson and J. S. Langer, Phys. Rev. Lett. 62, 2632(1989).
\bibitem{bak1} P. Bak, C. Tang and K. Wiesenfeld, Phys. Rev. Lett. 59, 381(1987), Phys. Rev. A38, 364(1988).
\bibitem{bak2} C. Tang and P. Bak, Phys. Rev. Lett. 60, 2347(1988); J. Stat. Phys. 51, 7977(1988).
\bibitem{burridge} R. Burridge and L. Knopoff, Bull. Seismol. Soc. Amer.57, 341(1967).
\bibitem{langer} J. S. Langer and C. Tang, Phys. Rev. Lett. 67, 1043(1991).
\bibitem{richter} B. Gutenberg and C. F. Richter, Ann. Geophys.9, 1(1956).
\bibitem{carlson3} J. M.Carlson and J. S. Langer, Phys. Rev. A40, 6470(1989).
\bibitem{souza} M. de Souza Vieira, Phys. Rev. Lett. 82, 201(1999).
\bibitem{meyers} C. R. Myers and J. S. Langer, Phys. Rev. E47, 3048(1993).
\bibitem{diamant} H. Diamant and D. Andelman, Phys. Rev. E61, 6740(2000).
\bibitem{kac} M. Kac and E. Helfand, J. Math. Phys. 4, 1078(1963); G. A. Baker Jr., Phys. Rev. 130, 1406(1963).
\bibitem{lieb} E. H. Lieb and D. C. Mattis, \emph{Mathematical Physics in One Dimension} (Academic Press, New York, 1966)chap.1.
\bibitem{sarker} S. K. Sarker and J. A. Krumhansl, Phys. Rev. B23, 2374(1981).
\bibitem{dik} A. M. Dikand\'e and T. C. Kofan\'e, Physica D83, 450(1995);  A. M. Dikand\'e, Phys. Lett. A220, 335(1996).
\end{thebibliography}
\end{document}